\documentstyle[emulateapj,psfig,apjfonts]{article}

\begin{document}

\title{A Strategy for Finding Near Earth Objects with the SDSS Telescope}

\author{Sean N. Raymond\altaffilmark{1,2}, 
Oliver J. Fraser\altaffilmark{2},
Arti Garg\altaffilmark{2},
Suzanne L. Hawley\altaffilmark{2}, 
Robert Jedicke\altaffilmark{3},
Gajus Miknaitis\altaffilmark{2},
Thomas Quinn\altaffilmark{2},
Constance M. Rockosi\altaffilmark{2},
Christopher W. Stubbs\altaffilmark{2},
Scott F. Anderson\altaffilmark{2},
Craig J. Hogan\altaffilmark{2},
\v{Z}eljko Ivezi\'{c}\altaffilmark{4,5},
Robert H. Lupton\altaffilmark{4},
Andrew A. West\altaffilmark{2}
Howard Brewington\altaffilmark{6},
J. Brinkmann\altaffilmark{6},
Michael Harvanek\altaffilmark{6},
Scot J. Kleinman\altaffilmark{6},
Jurek Krzesinski\altaffilmark{6,7},
Dan Long\altaffilmark{6},
Eric H. Neilsen\altaffilmark{8},
Peter R. Newman\altaffilmark{6},
Atsuko Nitta\altaffilmark{6},
Stephanie A. Snedden\altaffilmark{6}}

\altaffiltext{1}{Corresponding author: raymond@astro.washington.edu}
\altaffiltext{2}{Department of Astronomy, University of Washington, Box 351580,
Seattle, WA 98195}
\altaffiltext{3}{Institute for Astronomy, University of Hawaii, Honolulu, HI 96822}
\altaffiltext{4}{Princeton University, Princeton, NJ 08544}
\altaffiltext{5}{H.N. Russell Fellow, on leave from the University of Washington}
\altaffiltext{6}{Apache Point Observatory, P.O Box 59, Sunspot, NM 88349-0059}
\altaffiltext{7}{Mt. Suhora Observatory, Cracow Pedagogical University, ul. Podchorazych 2, 30-084 Cracow, Poland}
\altaffiltext{8}{Fermi National Accelerator Laboratory, P.O. Box 500, Batavia, IL 60510}

\begin{abstract}

We present a detailed observational strategy for finding Near Earth Objects
(NEOs) with the Sloan Digital Sky Survey (SDSS) telescope.  We investigate
strategies in normal, unbinned
mode as well as binning the CCDs 2$\times$2 or 3$\times$3, which affects the
sky coverage rate and the limiting apparent magnitude.  We present results
from 1 month, 3 year and 10 year simulations of such surveys.  For each
cadence and binning mode, we evaluate the possibility of achieving the
Spaceguard goal of detecting 90\% of 1 km NEOs (absolute magnitude H $\leq$ 18
for an albedo of 0.1). 
We find that an unbinned survey is most effective at detecting H $\leq$ 20
NEOs in our sample.  However, a 3$\times$3 binned survey reaches the
Spaceguard Goal after only seven years of operation.  As the proposed
large survey telescopes (PanStarss; LSST) are at least 5-10 years from
operation, an SDSS NEO survey could make a significant contribution to the
detection and photometric characterization of the NEO population.

\end{abstract}

key words:  solar system: general --- minor planets, asteroids --- surveys

\section{Introduction}

Collisions with Near Earth Objects (NEOs) pose a devastating threat to the
safety of the Earth.  (An NEO is an asteroid or comet with perihelion distance
$q \leq$ 1.3 AU and aphelion distance $Q \geq$ 0.983 AU.)
Current estimates of the global impact threat suggest that a collision
with an object of 1 km diameter is likely to cause global disaster in the
form of massive loss of food crops and tsunami-generated flooding,
while larger asteroid (10 km) impacts would likely result in near-global
extinction of the majority of advanced life forms
(Morrison, Spaceguard Survey report, 1992). Discovering and characterizing
large NEOs is therefore a high priority from a hazard viewpoint.
To address this hazard,
Congress and NASA developed the Spaceguard goal,\footnote{
http://impact.arc.nasa.gov/reports/spaceguard/} first
stated in 1995 and implemented in 1998, to discover 90\% of NEOs
larger than 1~km in diameter within ten years (Shoemaker, NEO Survey
Workgroup report, 1995).  The Sloan Digital Sky Survey (SDSS)
2.5m telescope and camera system (York et al. 2000) are uniquely positioned to
contribute to this initiative, with an ecliptic imaging survey that
is complementary to current NEO programs.

Operating in ``drift scan (time delay and integrate)'' mode, SDSS has been shown to be very effective at
detecting and characterizing Solar System objects (Ivezi\'c et al. 2001, 2002;
Juri\'c et al. 2002) using the movement
of the objects between the five filters in a single scan to discriminate
them from the stationary background.  Most of the asteroids observed in this
way
are new detections, because moving objects can be 
found over a wider field at fainter magnitudes than the completeness
limits of currently available asteroid catalogs.  However, the main
survey strategy results in only one imaging scan for a given region
of sky, and therefore orbits have not been obtained for these objects.

Matching the SDSS moving objects catalog against asteroids with known orbits 
has demonstrated that asteroid colors in the SDSS passbands are
tightly correlated with orbital parameters (Ivezi\'c et al. 2002).
This work has contributed significantly to the study 
of asteroid families, which are thought to be the remnants of disrupted 
larger bodies, and therefore to the study of the origin of the Solar System.
The asteroid colors are also important for assessing
the impact hazard, as there is a factor of 4
difference in the albedos of different asteroid taxonomic classes, and the
uncertainty in albedo translates directly into an uncertainty in
size and therefore potential hazard.  SDSS multicolor photometry 
will greatly improve our ability to classify asteroids and to determine
their spectral reflectivity.  While other NEO projects
have restricted themselves to a single wide-band filter or no filter
at all, the SDSS system produces excellent 
photometry in 5 colors.

Jedicke et al (2003) simulated the discovery of NEOs as a function of time, in
order to assess the current prospects for reaching the Spaceguard Goal of
identifying 90\% of NEOs larger than 1 km by 2008.
They assume an observing cadence typical of LINEAR (LIncoln Near Earth
Asteroid Research -- the most prolific 
NEO survey that is currently operational), and 
account for actual observing conditions by assuming only nighttime observing,
avoiding bright time and avoiding the Galactic Plane.  
They find that the LINEAR survey cannot meet the Spaceguard
goal.  Proposed surveys with limiting magnitudes $\sim$ 24 (LSST, Pan-Starrs)
are not likely to be operational for 5-10 years.  The only prospect for 
achieving the Spaceguard goal is therefore a dedicated survey in the V=21.5 
magnitude range.  As we will show in the following sections, the V=21.5
simulation corresponds closely to an NEO survey using the SDSS 2.5m
telescope and imaging system.  

In this paper, we simulate the detection of NEOs with the SDSS telescope using
a detailed cadence and observing strategy, which attempts to
maximize the detection and orbital characterization of NEOs.  We place
the results of these simulations in the context of the predictions of Jedicke
et al. (2003) to assess the possible contribution of an SDSS NEO survey toward
achieving the Spaceguard goal.  We also perform 10 year simulations of
NEO detection to evaluate in more detail the potential of an SDSS NEO survey.

\section{The SDSS camera and telescope}

The SDSS camera system (Gunn et al.  1998) consists of six columns of
five photometric CCDs, with a CCD for each of the five SDSS filters
(u, g, r, i, z) in each column.  The system is designed to image the sky by
driftscanning along great circles.  In this mode, the telescope is
driven in three axes so that the sky moves past the field of view
along one dimension of the CCDs.  The camera shifts the accumulating
image along the same dimension at the same rate, and the data is read
out when it reaches the edge of each CCD.  In this way, a continuous
image is built up in the scan direction.

The field of view of the camera is 2.5 degrees, each column of CCDs
being 13.5 arc minutes wide.  Two interleaved scans cover a given area
completely, with 10\% overlap.  The apparent magnitude limits for which the
SDSS system is 50\% complete are 22.5 (u filter), 23.2 (g), 22.6 (r), 21.9 (i)
and 20.8 (z).

Normally the SDSS driftscans at the sidereal rate, so that an object
is first detected by the r filter, and subsequently by the i, u, z and
g filters at intervals of 72 seconds.  Moving objects can be detected
by their apparent motion during the 5 minute time baseline between the
r and g filters, as demonstrated in Ivezi\'c et al. (2001), who
detected moving objects with apparent motion between 0.025 and 1
degrees per day.  Most asteroids are detected in the g, r and i
filters.

The SDSS telescope and camera can be used to driftscan faster than the
sidereal rate.  The camera has a minimum readout time per pixel very
close to sidereal, set by the design of the CCD controller and the charge
transfer characteristics of the detectors themselves.  Therefore,
faster scanning is done by binning the CCDs, thereby reducing the readout
time of the detectors.   In order to avoid smearing the images, the rate at
which the image of a star moves across the CCDs must be equal to the readout
rate of the CCDs.  The CCD readout time does not scale linearly with the
binning: in 2$\times$2 binned mode the minimum readout time is
one third of its unbinned value, and in 3$\times$3 binned mode it is roughly
one fourth.  Therefore, in 2$\times$2 binned mode, the telescope can cover
three times as much area as in unbinned mode.  The apparent magnitude limit
decreases by roughly one magnitude to 22.1 (g), 21.6 (r) and 21.0 (i).
The limits on
apparent motion scale roughly with the scan rate (modulo the astrometric
precision, which also varies with scan rate), and are roughly 0.1 --
1.5 deg day$^{-1}$ in 2$\times$2 binned mode.

Note that the magnitude limits given are 5$\sigma$ detections (S/N $\geq$ 5).
More relevant to the detection of moving objects is the apparent magnitude for
which some fraction of moving objects is detected.  Jedicke et al. (2003)
define V$_{50\%}$, the V magnitude for which 50\% of moving objects are
detected.  Ivezi\'c et al. (2001) and Juri\'c et al. (2002) demonstrated that
$\sim$ 90\% of moving objects are detected in unbinned mode to 21.5 magnitudes
in r.  The main difficulty in detecting fainter moving objects is the large
number of false positives.  We propose to detect $>$50\% of moving
objects to the 5$\sigma$ magnitude limit.  It should be much easier to remove
false positives with a dedicated survey such as we are proposing, because any
given patch of sky will be observed multiple times.  

\section{NEO Model Population}

Our simulations are based on the theoretical NEO population
of Bottke et al (2002).  Their model assumes that the NEO population is in a
steady state, being populated by small bodies from five source regions: (i) the
$\nu_6$ secular resonance in the main asteroid belt (37\% $\pm$ 8\%), (ii) the
3:1 Jupiter mean motion
resonance at 2.5 AU (25\% $\pm$ 3\%), (iii) the Intermediate Mars Crossing
region (23\% $\pm$ 8\%), (iv) the outer main asteroid belt (8\% $\pm$ 1\%),
and (v) Jupiter Family Comets (JFCs) (6\% $\pm$ 4\%), low-inclination comets
with origins in the Kuiper Belt (Levison \& Duncan, 1997).  The only population of
solar system objects not included in this model which poses a threat to the
Earth are the Nearly Isotropic Comets (NICs), which originate in the Oort Cloud
and make infrequent passages to the inner solar system.

\begin{figure}[p]
\centerline{\psfig{figure=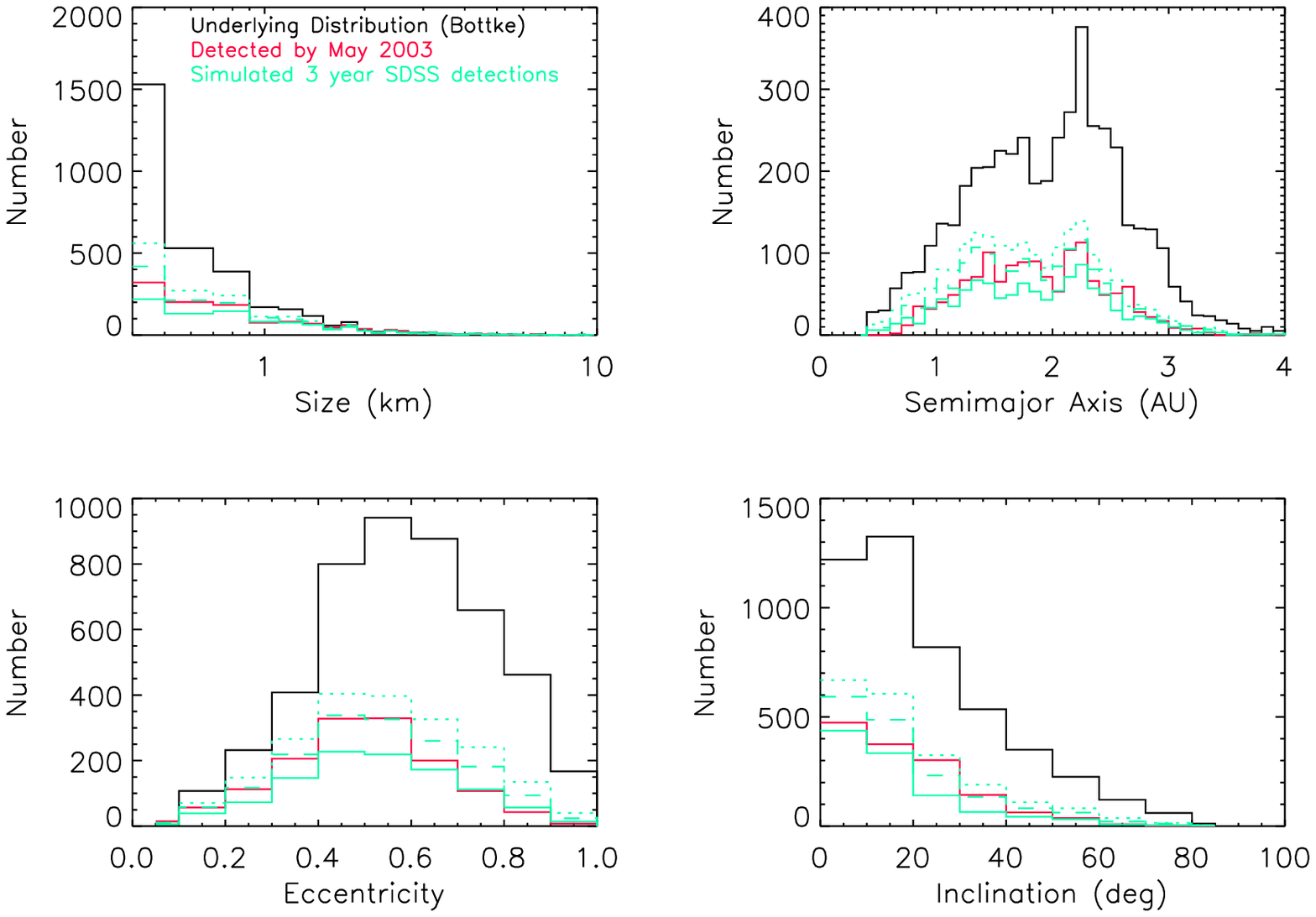,height=8cm}}
\caption{The orbital parameter and size distributions of NEOs with H $\le$ 20.
The red line
indicates known NEOs as of May 2003.  The black line is the Bottke et al
(2002) model, which we take as the underlying distribution.  The number of
currently known NEOs in this absolute magnitude range is 1405, where the Bottke
model predicts 4668, including 961 larger than 1 km.  The green lines
are the NEOs detected by a 3 year simulation using our NEO
strategy with the SDSS system.  The dotted green line
indicates NEOs detected once in our simulation (1933 objects), 
the dashed line indicates NEOs with multiple detections (1533 objects), and 
the solid green line indicates NEOs for which accurate orbits may be 
determined (3 detections in 10 days; 1023 objects) using only the 2.5m
data.  A followup program using the 3.5m and/or one of the smaller APO
telescopes could provide orbits for the remaining NEOs that were
detected only once or twice.  Note that our model includes objects as small as
400 m, but the top left panel begins at 600 m to avoid overshadowing $\geq$ 1
km NEOs. } 
\label{fig:s1}
\end{figure}

The sample distribution we used (obtained from W. Bottke, private
communication) contains 4668 NEOs with absolute magnitude H $\le$ 20
(diameter D $\geq$ 400m assuming an albedo of 0.1), including 961 NEOs larger
than 1 km (H $<$ 18).  The population has orbital eccentricities centered
near e=0.6, orbital
inclinations peaking near 10-20 degrees but with a significant tail to
higher inclination, and semimajor axes ranging from 0.5 AU to past 4 AU with a
broad peak near 2 AU.  The distributions of the NEO model population
in each of these parameters are shown with the black lines in Figure~\ref{fig:s1}.
The red lines indicate the distributions of the currently known NEO population
(from the Minor Planet Center database, as of May, 2003).  

It is clear from Figure~\ref{fig:s1} that a
significant number of asteroids with d $>$ 1 km remain to be found.

\section{Simulation Parameters}

We use the Bottke NEO model
as an underlying population, and sample it using the SDSS imaging
system at various binning, cadence, position, etc. to optimize the
number of NEO detections that lead to orbital solutions.  The simulations
are guided by the following information:

\noindent{\bf Weather:} 
Weather statistics for Apache Point Observatory (APO) were compiled from 
two sources.  Data from the Astronomy Research Consortium (ARC) 3.5m telescope
averaged over several years show that
approximately 35\% of the available observing time is lost to weather, 
15\% of the time is photometric, and 50\% is ``spectroscopic'', i.e.
the dome is open and data are being obtained, but the conditions are
not photometric.  SDSS spectroscopic observing logs, scaled to
include bright time, indicate that $\sim$ 45\% of the available
time is used for spectroscopy.  This NEO survey was intended to be conducted
in parallel with other science programs requiring photometric conditions.  We
therefore adopt the ``spectroscopic'' time for our simulations.  This amounts
to an average of 15 nights per month.  If this survey
were the sole program using
the SDSS telescope, it would average 19-20 usable nights per month, and would
improve the results in the upcoming sections.

\begin{figure}[p]
\centerline{\psfig{figure=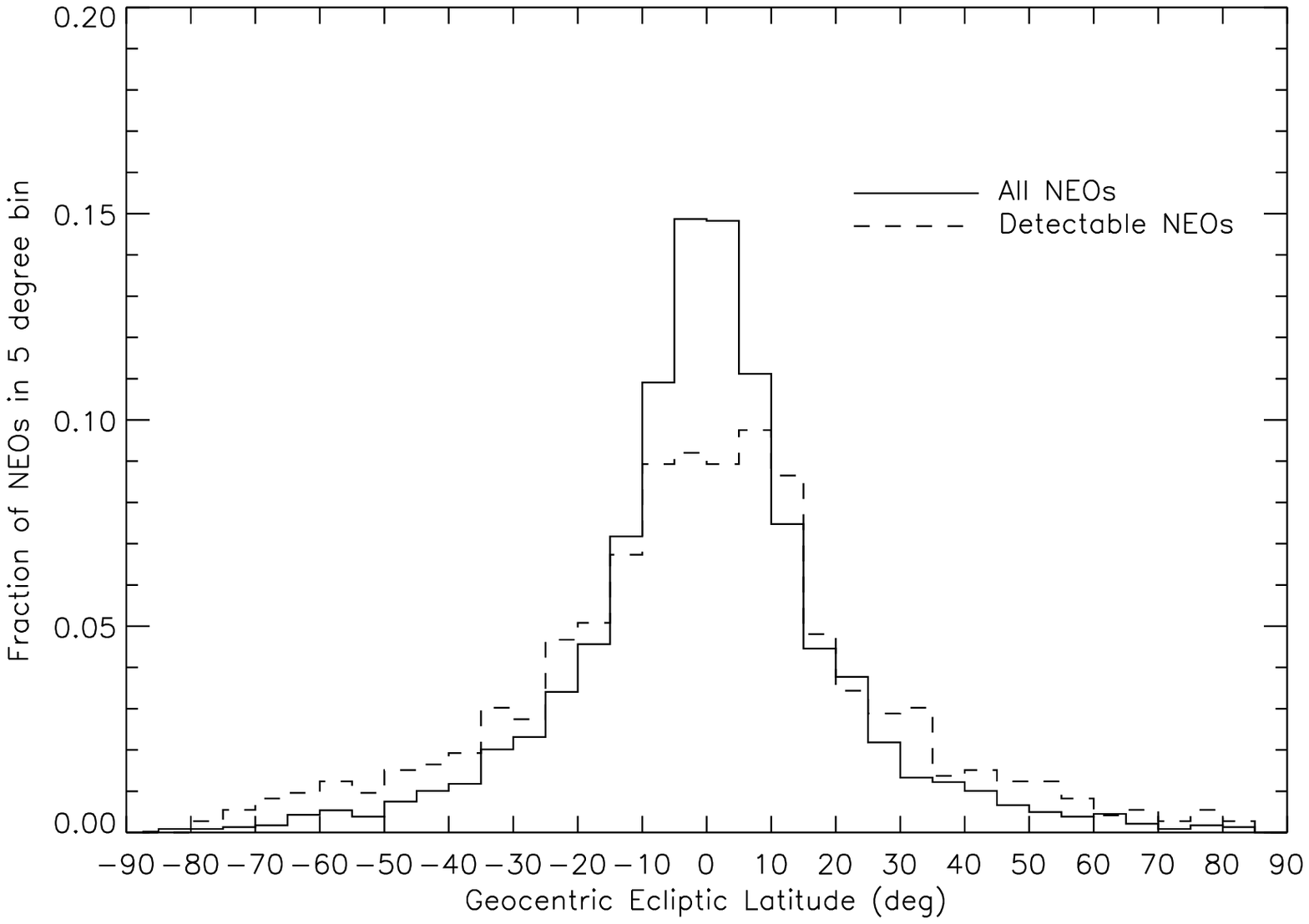,height=5cm}}
\centerline{\psfig{figure=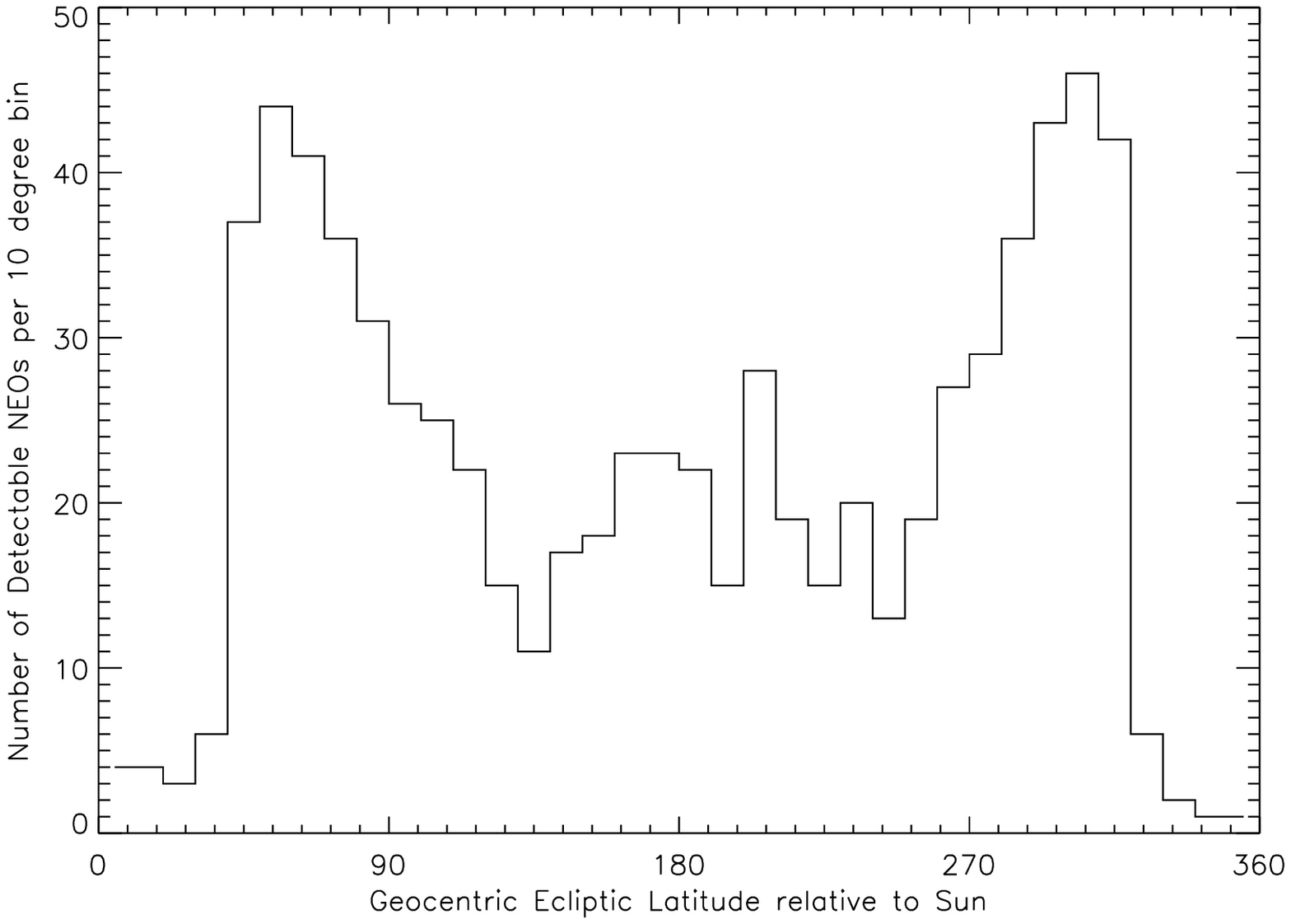,height=5cm}}
\caption{The instantaneous distribution of NEOs from the Bottke et al (2003)
model, as a function of geocentric ecliptic latitude (top) and longitude
(bottom). The bottom panel shows the increased detectability of the NEO
population at the sweet spots and opposition.  See text for details.}
\label{fig:s2}
\end{figure}

\noindent{\bf Ecliptic Latitude:}
Figure~\ref{fig:s2} (top panel) shows that the Bottke model is roughly
approximated by a Gaussian distribution with a FWHM of about 10 degrees.
The bright, nearby, ``detectable'' population (r $<$ 21.5) has a broader
distribution, similar to a Gaussian with a FWHM of about 20 degrees.
This suggests that an NEO survey should not concentrate entirely on the ecliptic
equator, but should extend in latitude on each side of the equator.  However,
the marginal gains decrease at latitudes past $\pm$20-30$^{\circ}$.  Note that
at any given time, roughly 1000 of the 4668 NEOs in our sample are
detectable.  Thus, any survey must operate for several years in order to detect a sizeable
fraction of NEOs, with a survey length that correlates inversely with limiting
magnitude.  Note that survey length does not scale linearly with
detectability, as it takes far longer to detect the second 50\% of an NEO
population than the first 50\%.  We return to this issue in sections 7 and 8. 

\noindent{\bf Solar Elongation:}
Figure~\ref{fig:s2} (bottom panel) plots the detectability of NEOs as a
function of solar elongation in the ecliptic plane (i.e. geocentric ecliptic
longitude at zero geocentric latitude, as measured on March 21).  A solar
elongation of 180 degrees corresponds to opposition, while 0 (and 360) degrees
is the position of the Sun.  Competing effects conspire
to give the three peaks in the diagram.  Objects at opposition
are completely illuminated by the Sun, and therefore have brighter apparent
magnitudes, and also have observable velocities favorable for detection.  
However, the volume inhabited by the NEO population which is encompassed
within the solid angle of the detector is a minimum at opposition, which
reduces the density of objects on the sky.  

The peaks at solar elongations of roughly $\pm$ 60 degrees (i.e. near
60 degrees and 300 degrees; called the ``sweet spots'') also result from
competing effects.  The density of NEOs on the sky is highest in the
direction of the Sun, as most NEOs are farther away from the Earth, and a
larger volume is
being sampled in that direction than away from the Sun, at opposition.
However, observing within about 45 degrees of the Sun is very difficult,
especially for ground-based telescopes.
NEOs which lie between the Earth and Sun will show large phases, and 
therefore be faint.  The holes in the distribution near elongations of 120 and
240 degrees are the places in the orbits
where the velocities are not favorable for detection (i.e. the apparent
velocities drop below 0.1 deg day$^{-1}$).  The sweet spots arise as a
balance between these effects, and are a prime location for detecting and
characterizing NEOs.

The above reasoning argues for an observing strategy that targets specific
solar elongations on a given night, that is breaking the night up
into several short scans -- see Figure~\ref{fig:s3} -- rather than obtaining a
single scan that covers the entire observable ecliptic plane.

\section{Observing Strategy}

We investigated unbinned (1$\times$1), binned (2$\times$2) and binned (3$\times$3) scans,
extending to $\pm$10-40 degrees in ecliptic latitude, and
with various ecliptic longitude sampling.  Our results
are as follows:

\noindent{\bf Binned vs. Unbinned:}  As mentioned above, unbinned observations
scan the sky at the sidereal rate,
2$\times$2 binned observations scan the sky three times faster, and 3$\times$3
binned observations four times faster.  The approximate limiting magnitudes for
unbinned, 2$\times$2 and 3$\times$3 binned scans in the r band are
22.5, 21.5 and 21.1, respectively.  The limits on apparent velocity are
determined by the time between exposures in the r and g filters, and are 
roughly 0.025-1 deg day$^{-1}$, 0.1-3 deg day$^{-1}$ and 0.15 - 4 deg day$^{-1}$ for the three
cases.  This trade-off
between sky coverage and depth of coverage is investigated in the upcoming
sections. 

\noindent{\bf Calculation of Observables:}
We used the JPL ``Horizons'' program 
(Giorgini et al 1996) to generate ephemerides for
each object in our NEO population each night for a period of three 
years.  From our
knowledge of the position of each object in space we calculate the relevant
parameters: phase angle, solar elongation, heliocentric and
geocentric distances.  We calculate the size distribution of NEOs, assuming all
NEOs have an albedo of 0.1.  The apparent magnitude is calculated following
the method of Bowell et al (1989), assuming a
slope parameter G of 0.23, which is typical for S type asteroids, and a
reasonable, albeit slightly higher than average, value for NEOs (Morbidelli et
al 2002).  The result of these calculations is a (large) data table
that gives us all the observable quantities for each NEO in the population
at each time throughout the three year simulation.

\begin{figure}[p]
\centerline{\psfig{figure=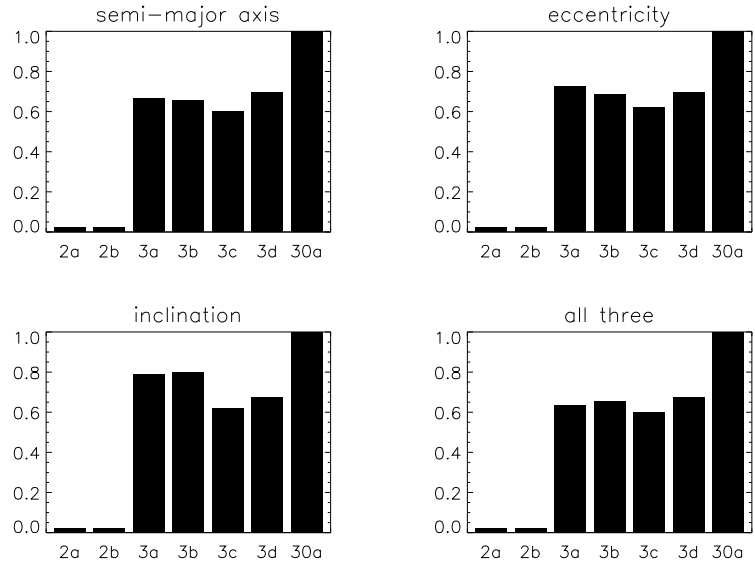,height=8cm}}
\caption{Results of fitting orbits with various sample cadences are
illustrated.  The cadences are:
\ \ \ 2a = days 1 and 2 
\ \ \ 2b = 1,3
\ \ \ 3a = 1,3,10
\ \ \ 3b = 1,5,10
\ \ \ 3c = 1,10,20
\ \ \ 3d = 1,15,30
\ \ \ 30a = all 30 days.  The histograms show the fraction of objects
in a one month simulation whose
primary orbital parameters (eccentricity, semi-major axis, inclination,
and the combination of all three of these parameters) can be
reproduced within 1\% of the input values, for each of these cadences.
}
\label{fig:bargraph}
\end{figure}

\noindent{\bf Orbits:} 
Previous efforts (Bowell et al. 2002) indicate that 3-4 observations
spaced over a few up to 20 days will provide a good fit to a
typical NEO orbit.  To test our ability to recover orbits, we use the output
of a given month of simulation as input to an orbit finding
program (ORBFIT- Milani 1999).  We used the detections on both the g and r
chips (hence an instantaneous velocity vector) in the orbit fitting.
Figure~\ref{fig:bargraph} shows that orbits are recovered at the 65-70\% level
if three observations are available, and that the spacing of the three
observations is not critical for any of the individual parameters.  As this
represents only one of a number of methods for determining orbits, we consider
this value to be a lower limit on the fraction of orbits accurately obtained.  

This requirement for orbital determination, together with the weather statistics, 
constrains the frequency and number of revisits to a given field, and
therefore the cadence.

\noindent{\bf Auxiliary Data:}
The possibility of obtaining auxiliary
followup data on other telescopes would greatly increase the number of
objects found with good orbits, allowing translation along the
green curves in Figure~\ref{fig:s1} from the dotted line (1 detection) to
the dashed line (2 detections) to the solid line (3 detections,
and well determined orbit), giving roughly a factor of two increase
in NEOs with orbits over those that could be found using the
SDSS system alone.  We note
that detection of the objects is the primary requirement, and
possible systematic differences in photometric calibration are
not a major concern.

\noindent{\bf Cadence:}
In order to maximize the number of NEOs for which accurate orbits are
determined, we devised a strategy whereby we observe
two contiguous (in latitude) stripes (each stripe consists
of two interleaved strips) on alternating nights until
both have been observed three times.  We then offset in latitude
and observe two new stripes, again alternating.  Experience
with the simulations suggests that approximately 5-6 stripes
can be observed in a given month.
We find that since most of
the ecliptic longitude imaged in a given month is still available
in the following month (basically the opposition and eastern
peaks in the solar elongation plot, Figure~\ref{fig:s2}, are 
still available), we can increase
the total area observed by adopting a bi-monthly strategy,
with one month investigating latitudes north
of the ecliptic, and the next month south.  This strategy balances the
competing requirements to obtain several (of order 3) observations of each
object and to cover as much area as possible.  Figure~\ref{fig:s3}
is a representation of the first three stripes in the northern direction.
Note that, although the stripes are shown as scans parallel to the
ecliptic equator, these are great circles on the sky, so there is
a slight curvature that is not depicted on the plot.

\noindent{\bf Airmass Constraints:}
The SDSS telescope is located at Apache Point Observatory (APO) in New Mexico,
at a latitude of 32.8$^{\circ}$ North.  The 
minimum zenith angle of the ecliptic at local midnight therefore oscillates
between 9.3$^{\circ}$ on the (northern) winter solstice and 57.3$^{\circ}$ on
the summer solstice.  An optimized NEO
survey may want to avoid observations at airmasses larger than 2 (zenith
angles $>$ 60$^{\circ}$).  In that case, we would not want to observe south of
the ecliptic in the summer months.  A strategy could be devised whereby scans
on consecutive summer months would focus solely on northern ecliptic
latitudes.  Alternatively, the ecliptic latitude about which monthly scans are
centered could shift to 10-20 degrees north during the summer months.  This
issue has not been addressed to date in our cadence, but the effect
on the realism of our simulations is not large, particularly since the length
of the observable night is at a minimum when the ecliptic is most difficult to
observe, in the summer.  

\section{Results}

\begin{figure}[p]
\centerline{\psfig{figure=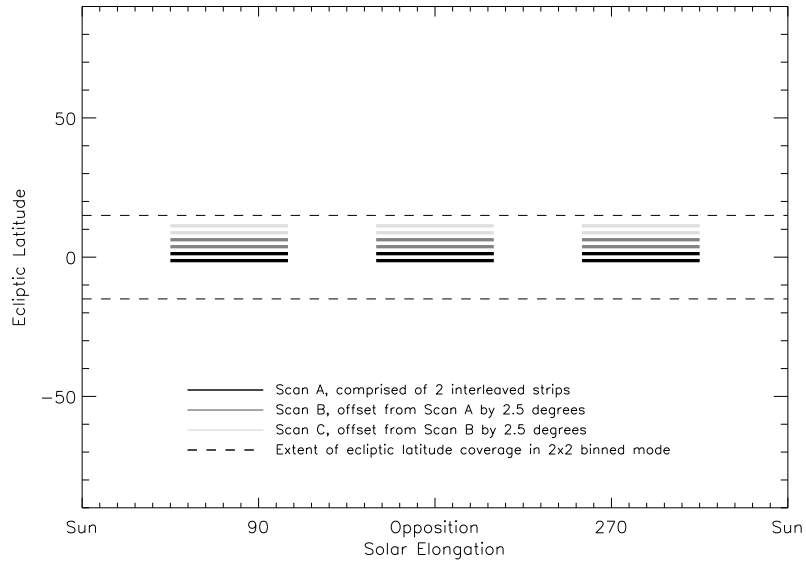,height=8cm}}
\caption{Our NEO cadence for 2$\times$2 binning mode, focusing on the
sweet spots and opposition.  Each night consists of six 60-degree scans
($\pm$ 1.25 hours each), to fill a stripe (e.g. A).  We observe the same
stripe on consecutive nights, and shift in ecliptic latitude after a given
stripe has been observed 3 times.  Starting on the ecliptic, we shift to
northern and southern latitudes in consecutive months, then repeat.  A
cadence for the 15 observing nights in one month is A, A, B, B, A,
B, C, C, D, D, C, D, E, E, F, where each lettered stripe is offset from the
previous one by 2.5 degrees.  Stripes A, B and C are illustrated
in the figure (not to scale).  The stripes are great circles and will exhibit
slight curvature in this coordinate system (not shown).} 
\label{fig:s3}
\end{figure}

\begin{figure}[p]
\centerline{\psfig{figure=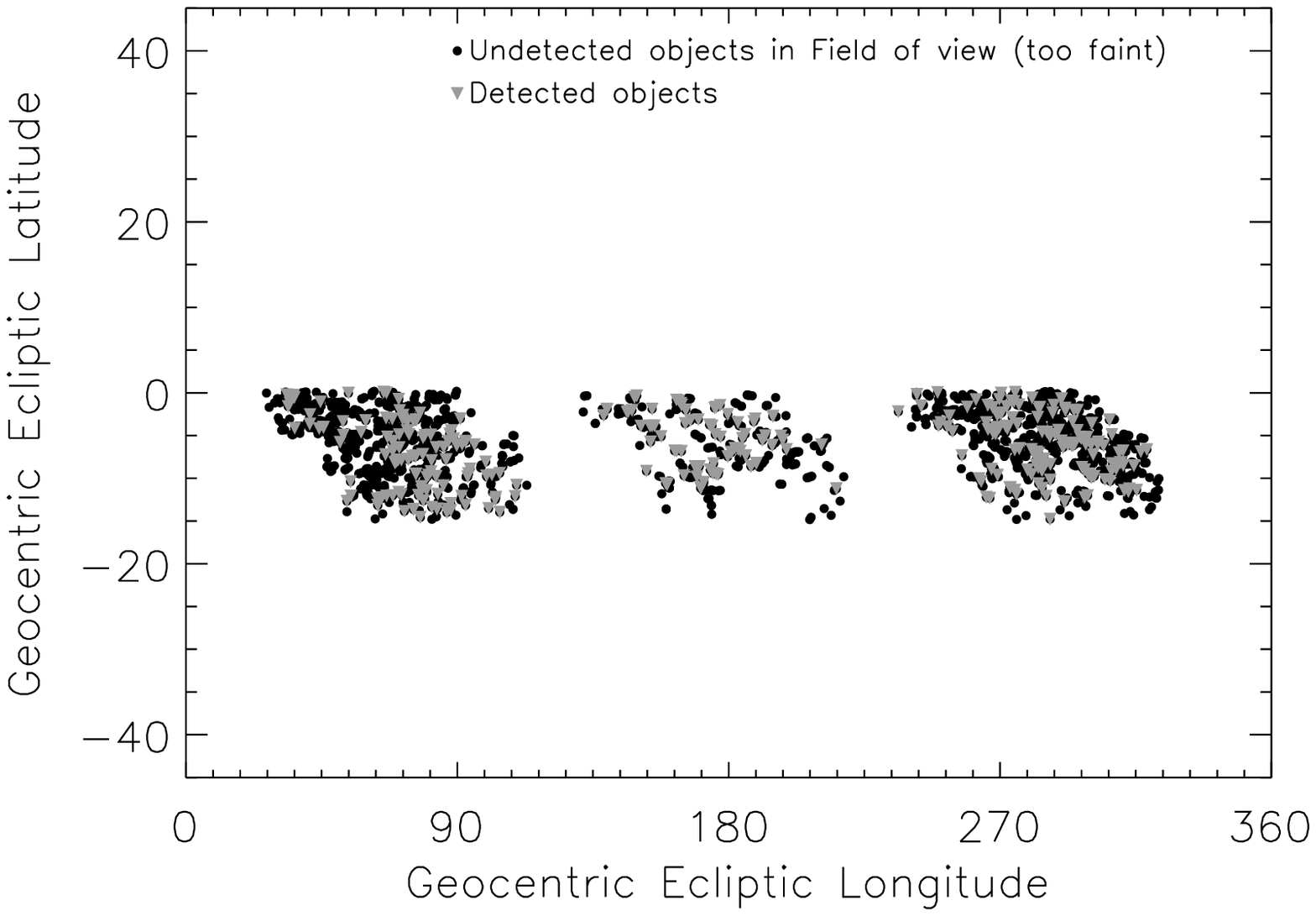,height=5cm}}
\centerline{\psfig{figure=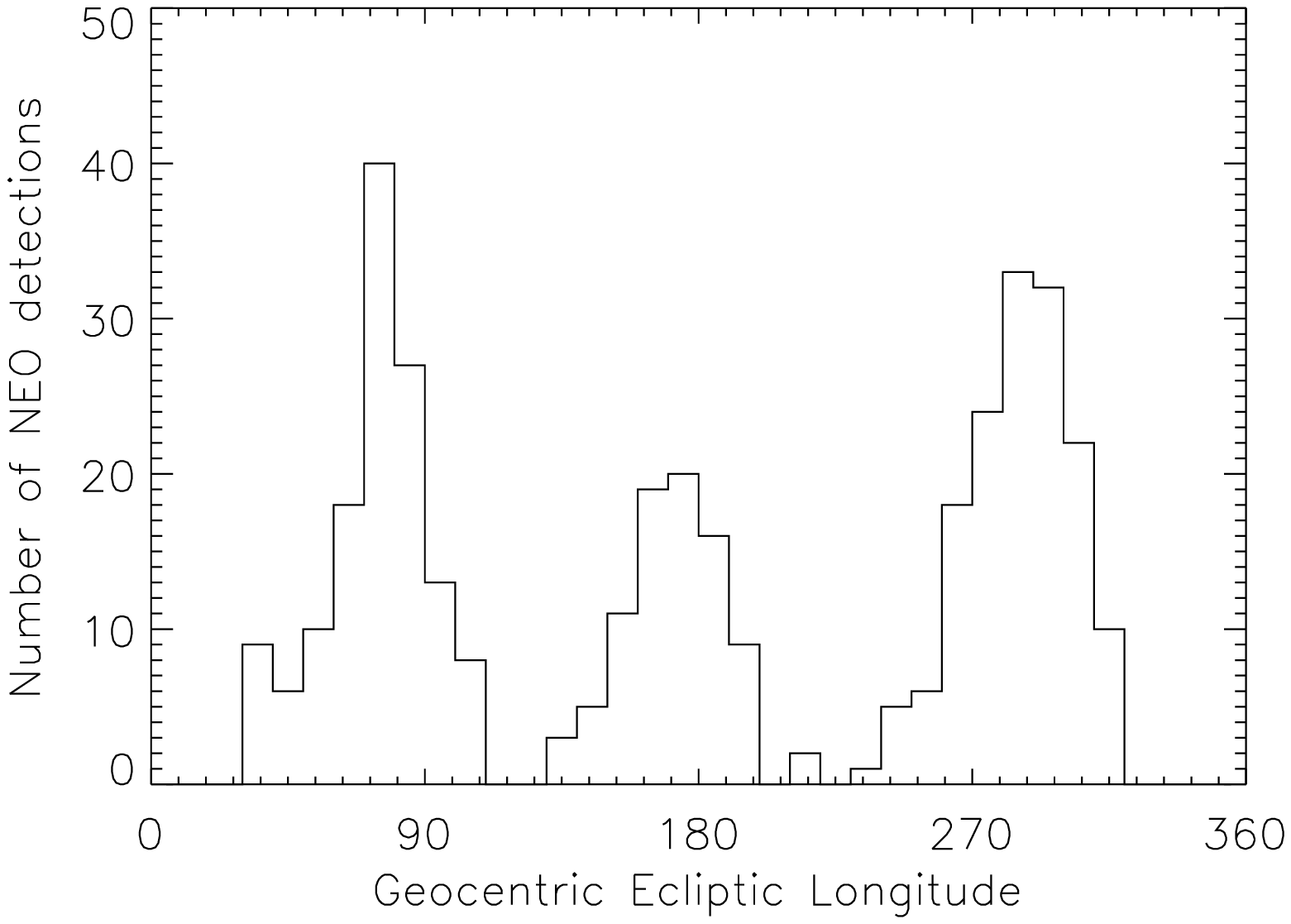,height=5cm}}
\caption{The top panel shows detections (grey dots) in a 1 month 
simulation as a function of ecliptic latitude and geocentric ecliptic
longitude.  The month of March was used, so that opposition corresponds
to longitude of 180 degrees at the vernal equinox.  The black dots 
indicate objects which fell inside the camera's field of view, but were 
too faint for a 5$\sigma$ detection.  
The parallelogram shape results
from the movement of opposition and the sweet spots in the ecliptic coordinate 
system by 30 degrees through the month.
The histogram on the bottom is made up solely of the detected objects, and
shows the sweet spots expected from Figure~\ref{fig:s2}. }
\label{fig:s4}
\end{figure}

\begin{figure}[hp]
\centerline{
\psfig{figure=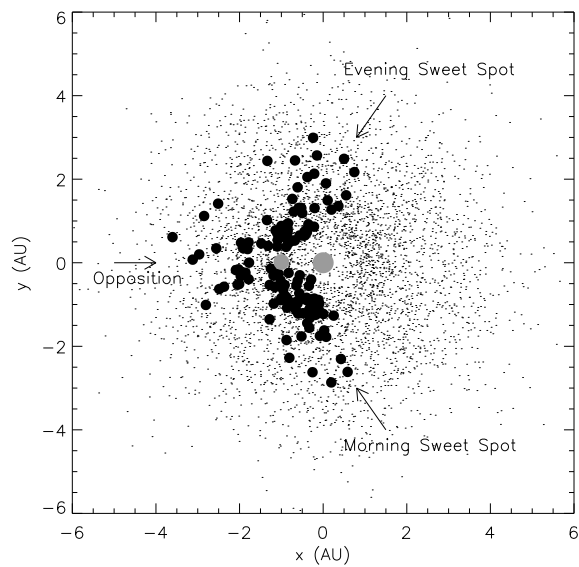,height=8cm}}
\caption{Detections in the 1 month simulation from Figure~\ref{fig:s4}, in
x-y coordinates. The Sun is depicted in grey at the origin, and the Earth in grey
at the appropriate location in its orbit and a distance of 1 AU.  
The large dots represent NEO detections, and the small dots represent the
total instantaneous distribution of NEOs.  The sweet spots and opposition are
labeled, with the ``evening sweet spot'' being observed first after sunset.}
\label{fig:s5}
\end{figure}

Figures~\ref{fig:s4} and~\ref{fig:s5} show the results of a 1 month simulation,
using the cadence from Figure~\ref{fig:s3}, binned 2$\times$2.  The assumed 15
observable nights are randomly spread throughout the month, and we assume 8
hours of observing per night .  To account for
the presence of the
Moon, the apparent magnitude limit varies sinusoidally with a 28 day period
between r magnitudes of 20.5 and 21.5, which are conservative 5$\sigma$ 
detection limits for the binned data (see Section 2).
In this particular simulation, the 147 objects were detected at least once.  The
orbits of 58 NEOs were determined according to the
criterion of Bowell (3 observations in 10 days), and 102 objects were detected
at least twice.  Single and double observations can be submitted to the Minor
Planet Center for followup, and are valuable to the community and the
Spaceguard effort even without precise orbital elements.  However, note that
few current NEO surveys are as sensitive as SDSS, so this survey would be
largely responsible for its own followup observations. 

Our cadence is designed to observe each NEO 2-3 times, but $\sim$ 1/3 of the total
detected objects are seen only once.  There are two
ways in which an NEO can elude multiple detection.  If it is very close
to the Earth, and therefore moving very rapidly on the sky with a significant
component of its velocity perpendicular to the ecliptic, it will not be picked up
on successive scans of the same area.  In addition, the apparent magnitude
of NEOs in certain orbits can change by a significant amount within a
few days, through changes in their apparent phases and distances from Earth. 

Our 1 month scans are extended into a year-long strategy by alternating
excursions in latitude North and South of the ecliptic.  This cadence detects
10-20\% more NEOs in a 3 year simulation than one which scans the same area
each month, including roughly twice as many H $\leq$ 18 NEOs, the targets of
the Spaceguard Goal. 

\begin{figure}[p]
\centerline{\psfig{figure=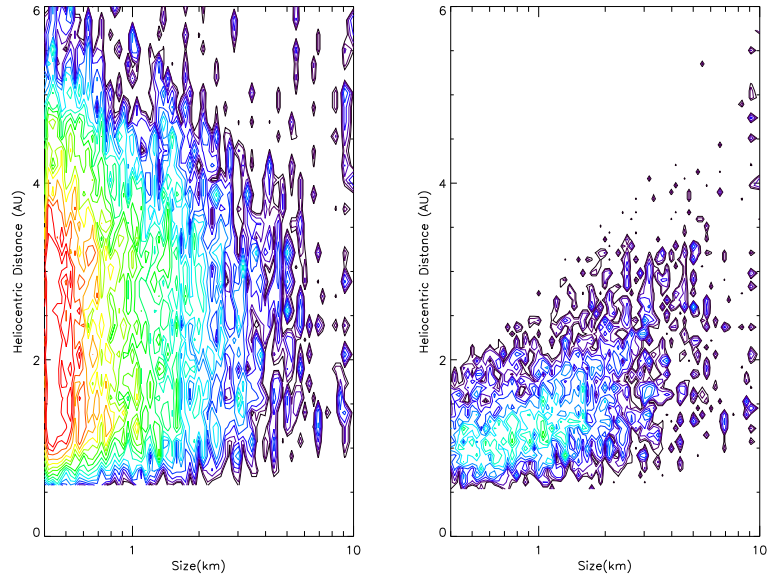,height=8cm,angle=90}}
\caption{The detection volume of NEOs in apparent magnitude - diameter space.
The left panel shows the underlying distribution of
detectable NEOs in a 3 year simulation, while the right panel shows the
ones detected with our proposed 3 year NEO survey operating in 2$\times$2 binned
mode.  The fraction of
detected objects increases with size, as their detection volume increases.
The detection volume is a function of the size, albedo, and phase of an
asteroid.  Our distribution assumes an albedo of 0.1 for all objects.  
The contours are logarithmically spaced for clarity.}
\label{fig:s6}
\end{figure}

\begin{figure}[ph]
\centerline{\psfig{figure=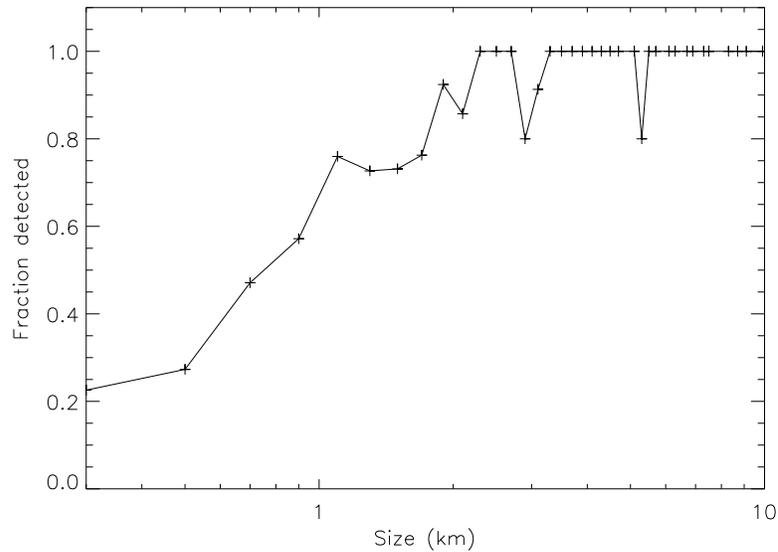,height=8cm}}
\caption{The fraction of NEOs in our field of view which were
detected as a function of size for the same 3 year simulation as 
in Figure~\ref{fig:s6}.  It is clearly much easier to detect larger objects
which have
correspondingly larger detection volumes.  Note that, of our sample of 4668
NEOs, 961 are larger than 1 km (assuming an albedo of 0.1) , and the number of
objects per bin is small past five km.}
\label{fig:s7}
\end{figure}

We have
run 3 year simulations using a variety of detailed cadences (i.e. the ordering
of scans A, B, C, etc. from Figure~\ref{fig:s3}).  Figures~\ref{fig:s6} and~\ref{fig:s7} 
show the results of a 3 year simulation operating in 2$\times$2 binned mode,
in which 1931 NEOs were detected at least once and 1552 at least
twice. Accurate orbits were derived for 994 objects.  Figure~\ref{fig:s6}
demonstrates that the detection volume of NEOs is a function of the NEO size.
Figure~\ref{fig:s7} shows the fraction of NEOs which are detected in a given
field of view, thereby quantifying the fraction of the detection volume
accessible for different-sized NEOs.

Table~\ref{table:sims} shows a comparison between unbinned, 2$\times$2 binned
and two 3$\times$3 binned
simulations.  Each simulation lasted three years, assumed 15 nights per
month of observing, and 8 hours per night.  In 1$\times$1 (unbinned) mode, 120
degrees of the ecliptic are scanned in a night, divided into three
pieces: scans along the two sweet spots and opposition.  The scans
cannot be interleaved in the same night, so it takes two full nights to cover
a 2.5$^{\circ}$ wide stripe on the ecliptic, thereby limiting the
range in ecliptic latitude that can be covered.  In 2$\times$2 binned mode,
the increase in scan rate means that two scans can be interleaved, each
consisting of three similar pieces which focus on the sweet spots and
opposition, as shown in figure~\ref{fig:s3}.  In 3$\times$3 binned mode, 480
degrees of the ecliptic are scanned in a night.  We have simulated two
different cadences in 3$\times$3 mode: (i) increasing
the length of scans to be nearly continuous in ecliptic longitude, keeping
roughly the same latitude coverage as in 2$\times$2 binned mode, and (ii)
concentrating on the sweet spots and opposition, and increasing coverage in
ecliptic latitude.  Case (ii) produced more NEO detections, as shown in
table~\ref{table:sims}. 

All three modes of operation result in roughly the same number
of 1 km NEOs detected.  Unbinned mode is the most proficient at single
detections, although a large number of these are smaller objects.  2$\times$2 binned
mode is the best at characterizing the orbits of NEOs, although only slightly
better than 3$\times$3(ii) mode, which detects slightly more 1 km NEOs.  Although a
vigorous followup program could attempt to re-observe the singly and doubly
detected NEOs, 2$\times$2 binned mode is the most efficient at self-followup, whereas
unbinned mode produces good orbits for only about one third of the NEOs it
detects.  

\section{Comparison with Jedicke et al. (2003) results}

As of May 2003, only 1405 NEOs with H $\le$ 20 (d $\ge$ 400m
assuming an albedo of 0.1) have been discovered (out of 2307 total NEOs
discovered).  In 3 years,
our NEO simulations show that an SDSS NEO survey
could come close to matching that number for objects with well-determined
orbits.  However, this number does not account for the
fraction of known NEOs rediscovered by SDSS.  

A detailed analysis by Jedicke et
al (2003) demonstrated that a deep survey with a limiting
magnitude of V $\sim$ 21.5 (similar to SDSS) would reach the Spaceguard goal 
10-20 years faster than the current leading NEO survey, LINEAR, which has a 
limiting magnitude of V$_{50\%}$ $\sim$ 19.  The Jedicke simulations assume 
a cadence which covers the entire observable sky each month,
which is not directly comparable to our 2$\times$2 binned strategy
that surveys only to ecliptic latitudes about $\pm$ 15 degrees.  

Operating in 3$\times$3 binned mode, SDSS has a limiting magnitude in the r
band of 21.1, which corresponds to V $\sim$ 21.4.  The sky coverage rate is
four times the sidereal rate, 60 degrees per hour.  Assuming 8 hours of
observing per night, SDSS covers 480 degrees of ecliptic longitude in a
given night.  To cover a given 2.5 degree-wide ``stripe'', two ``strips'' must
be interleaved.  Therefore, SDSS can cover a 2.5$^{\circ}\times$240$^{\circ}$
patch each night.  In 16-17 nights, SDSS can cover 10,000 deg$^2$ of sky,
roughly one quarter of the entire sky.  As the latitude distribution of
NEOs is not uniform (see figure~\ref{fig:s2}), a search program similar to
the one described above could cover the ecliptic to roughly $\pm$ 20$^{\circ}$ in
ecliptic latitude each month, a region in which $\gtrsim$ 80\% of NEOs are found.
Alternatively, a two month strategy could reach ecliptic latitudes of $\pm$
40$^{\circ}$, and $\gtrsim$ 95\% of NEOs.  This implies that SDSS, operating in 3$\times$3
binned mode, would be roughly equivalent to a V$_{50\%}$ = 21.5 survey from
Jedicke et al. (2003), and would be able to detect
90\% of NEOs larger than 1 km by about the year 2010 if such a program were
implemented starting in 2002.  However, this is a very indirect argument.  In
the following section we present the simulated results of 10 year NEO surveys
which include a realistic pre-detected population. 

\section{Long-Term Simulations}

\begin{figure}[ph]
\centerline{\psfig{figure=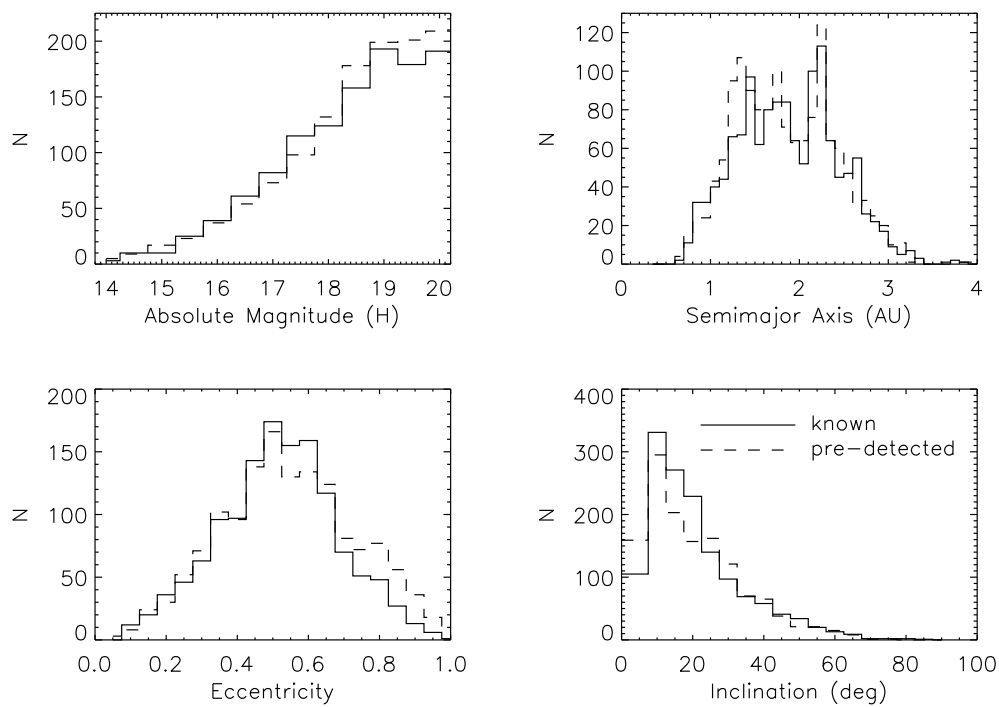,height=10cm}}
\caption{A comparison between our pre-detected population (dashed lines) and
the population of known NEOs (solid lines) as of Jan 1, 2003.  The agreement
between the samples indicates that this is a realistic representation of the
known NEO population.}
\label{fig:predet}
\end{figure}

To better evaluate the performance of an SDSS NEO survey, we have performed
long-term, 10 year simulations of NEO discovery using the different modes of
operation discussed in the previous sections.  We have included a realistic
pre-detected population of NEOs by running a ``pseudo-LINEAR'' simulation of
NEO detection as in Jedicke et al. (2003) on our sample of 4668 NEOs with H
$\leq$ 20.  The pre-detection simulation was run until it matched the 628
NEOs with H $\leq$ 18 known as of Jan 1, 2003.  At the end of this
simulation the total number of pre-detected NEOs was 1424, slightly more
than the 1333 H $\leq$ 20 NEOs known as of 1/1/2003 (data from the Minor
Planet Center).  The distributions of the known and pre-detected populations
are shown in Fig~\ref{fig:predet} and match up remarkably well, giving us
confidence that our pre-detected population is realistic.

\begin{figure}[ph]
\centerline{\psfig{figure=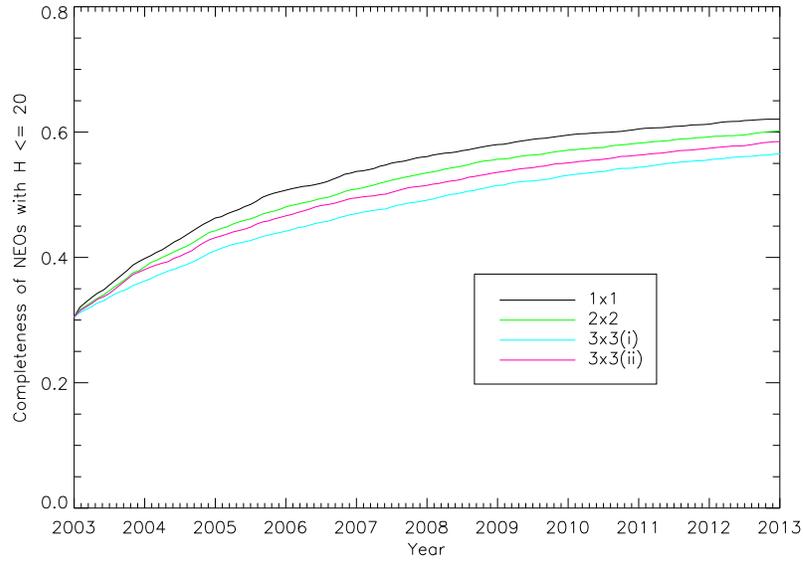,height=8cm}}
\caption{Completeness of all NEOs with H $\leq$ 20 in our sample as a function
of time, 
including the pre-detected population from Figure~\ref{fig:predet}, labeled
as in Table 1.  An NEO survey in unbinned (1$\times$1) mode is
the most efficient.  A 10 year unbinned survey increases the completeness of H
$\leq$ 20 NEOs from 30\% to 62\%.} 
\label{fig:comp}
\end{figure}

\begin{figure}[ph]
\centerline{\psfig{figure=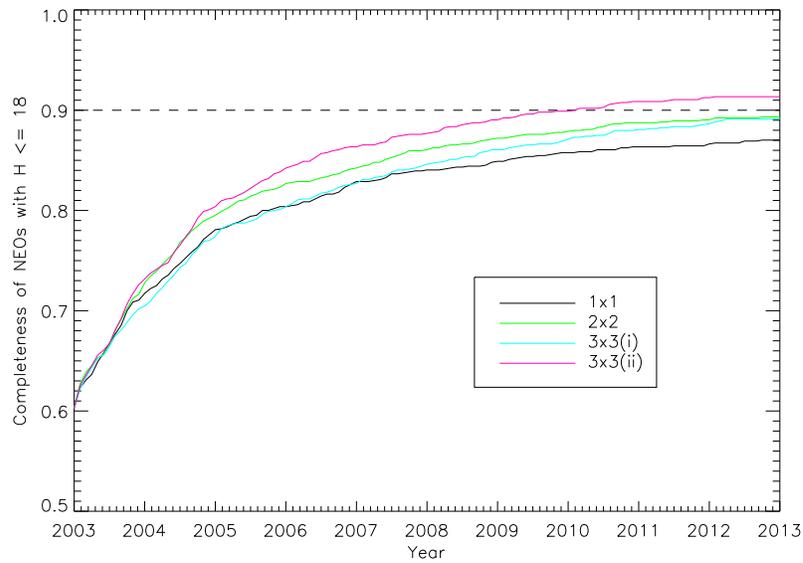,height=8cm}}
\caption{Completeness of NEOs with H $\leq$ 18 (d $\geq$ 1 km for an albedo of
0.1) as a function of time, labeled as in the previous figure.  The dashed
line represents a completeness of 90\%.  A survey in
3$\times$3(ii) binned mode detects larger NEOs the fastest, in contrast with
the previous figure which shows that an unbinned survey has the highest
detection rate for all NEOs with H $\leq$ 20.  A 3$\times$3(ii) binned survey
could have achieved the Spaceguard goal of detecting 90\% of H
$\leq$ 18 NEOs by early 2010, had it begun in January 2003. }
\label{fig:comp1k}
\end{figure}

The results of four 10 year simulations are shown in Figures~\ref{fig:comp}
and~\ref{fig:comp1k}, including NEOs with 2 or more detections in a span
of 10 days.  An unbinned survey, described in
Sections 5 and 6, detects the most NEOs in 10 years.
The completeness of the pre-detected population is 30\%, and
reaches 62\% after a 10 year survey in unbinned mode.  Completenesses are
calculated assuming the Bottke et al. (2003) model as an underlying NEO
population, as described in Section 3. 

If the goal of an NEO survey is to achieve the Spaceguard goal of detecting
90\% of NEOs with H $\leq$ 18, then Figure~\ref{fig:comp1k} shows that the
3$\times$3(ii) binned survey, described in Sections 5 and 6, is most
effective.  The completeness of the pre-detected population is 60\%, and
reaches 91\% by the end of a 10 year survey in 3$\times$3(ii) binned mode.
The 90\% mark is reached after seven years, in January 2010.  Note that this
is almost exactly the time predicted by Jedicke et al. (2003) for a survey
with limiting magnitude of V${50\%}$ = 21.5, and determined independently.

It is interesting that the optimal survey strategy depends on the goal of the
survey.  The balance between magnitude limit and sky coverage is such that a
3$\times$3(ii) binned survey is able to achieve the Spaceguard goal
the fastest, while an unbinned survey detects a significantly larger
fraction of the total NEO population.  A 2$\times$2 binned survey does
moderately well in both regimes, detecting a total of 89.3\% of NEOs after 10
years.

One can imagine a long-term NEO survey which begins operation in
3$\times$3(ii) binned mode, and transitions to unbinned mode after the
completion of the Spaceguard goal.  Such a hybrid survey would provide the
appropriate balance between survey depth and sky coverage, and adapt to the
current scientific need.

\section{Conclusions}

The detection of km-sized potentially hazardous asteroids is a high priority. 
We have demonstrated that the SDSS telescope and camera system has the ability
to detect and
characterize the orbits of NEOs at a fast rate.  We have
detailed four different cadences, which depend on the binning of
the CCDs, the spatial distribution of NEOs, and observational restrictions.  

In 3$\times$3 binned mode, the 5$\sigma$ detection limit corresponds
to V $\sim$ 21.4.  The 3$\times$3 areal coverage rate implies that an SDSS NEO
survey in that mode would be roughly equivalent to a V$_{50\%}$ = 21.5 NEO
survey from Jedicke et al (2003).  Their results, in turn, imply that
such a survey could reach the Spaceguard goal of detecting 90\% of NEOs with H
$\leq$ 18 by the year 2010, had such a survey begun in early 2002.

We have performed long term, 10 year simulations of SDSS NEO surveys with each
of our four cadences.  We find that an unbinned survey detects the largest
number of NEOs, 62\% of NEOs with H $\leq$ 20 in 10 years.  Alternatively, the
3$\times$3(ii) binned survey reaches the Spaceguard Goal
most quickly, in 2010 for a survey beginning in January, 2003.  This is very
close to the prediction of Jedicke et al. (2003).

The accurate, five-band photometry of the SDSS system would also be a huge
benefit to NEO science, through the composition and albedo (and therefore
potential
hazard) determination of NEOs.  A large side benefit to Solar System science
would be the serendipitous discovery and compositional determination of a
large number of small solar system bodies, main belt asteroids and Kuiper belt
objects.  An unbinned survey would be most beneficial in this
regard, due to its deeper exposures and frequent scanning of the ecliptic.

\section{Acknowledgments}

We gratefully acknowledge extensive discussions with our colleagues in the 
SDSS/APO community.  Our thanks to the SDSS Advisory Council and Management
Committee for supporting this effort to investigate an NEO survey, to Bill
Bottke for generating our NEO sample population, 
and to Ted Bowell and Al Harris for helpful discussions.  CH thanks NSF grant
AST-0098557 at the University of Washington for support.  SR and TQ are
grateful to the NASA Astrobiology Institute for support.

\newpage

\begin{deluxetable}{lcccc}
\tablecaption{Results of 3-year NEO Simulations}
\renewcommand{\arraystretch}{.6}
\tablehead{
  \colhead{ }&
  \colhead{1$\times$1\tablenotemark{a}}&
  \colhead{2$\times$2\tablenotemark{b}}&
  \colhead{3$\times$3(i)\tablenotemark{c}}&
  \colhead{3$\times$3(ii)\tablenotemark{d}}
}

\startdata

Limiting r magnitude & 22.5 & 21.5 & 21.1 & 21.1\\
Ecliptic Latitude Coverage & $\pm$10$^{\circ}$ & $\pm$15$^{\circ}$ & $\pm$20$^{\circ}$ & $\pm$30$^{\circ}$\\
NEOs with 1+ detections & 2184 & 1929 & 1671 & 1781\\
NEOs with 2+ detections & 1641 & 1533 & 1341 & 1439\\
NEOs with orbits  & 699 & 1023 & 566 & 991\\
NEOs ($>$ 1 km)\tablenotemark{e}  & 756 & 758 & 735 & 774\\
\enddata

\tablenotetext{a}{Unbinned survey, focusing on the sweet spots and opposition.}
\tablenotetext{b}{2$\times$2 binned survey, also focusing on the sweet spots and opposition.}
\tablenotetext{c}{3$\times$3 binned survey, extending to cover a wider region
around the sweet spots and opposition.}
\tablenotetext{d}{3$\times$3 binned survey, focusing on a narrower region
around the sweet spots and opposition, and extending to higher latitudes.}
\tablenotetext{e}{Number of $>$1 km NEOs detected at least once.}
\label{table:sims}
\end{deluxetable}
\clearpage

\end{document}